\documentstyle[11pt,newpasp,twoside,epsf]{article}
\newcommand{\nalph}{N/$\alpha$}
\newcommand{\alphH}{$\alpha$/H}

\newcommand{\lya}{Ly$\alpha$ }

\newcommand{\cm}[1]{\, {\rm cm^{#1}}}
\newcommand{\N}[1]{{N({\rm #1})}}

\def\edcomment#1{\iffalse\marginpar{\raggedright\sl#1\/}\else\relax\fi}
\reversemarginpar

\begin{document}
\title{The Physics of Extragalactic Gas: One Argument for a 
Next Generation Space Telescope}
 \author{Jason X. Prochaska}
\affil{UCO/Lick Observatory; UC Santa Cruz; Santa Cruz, CA 95064}

\begin{abstract}
Observations of the Galactic ISM have had tremendous impact on our 
understanding of the physics of galactic gas and the processes of
galaxy formation.  Similar observations at $z>2$ reveal the neutral
baryonic content of the universe, trace the evolution of metal
enrichment, shed light on process of nucleosynthesis and dust formation,
and yield precise measurements of galactic velocity fields.
Owing to the limitations of UV spectroscopy, however, researchers are
unable to examine galactic gas at $0 < z < 2$, an
epoch spanning $\approx 80\%$ of the current universe.  
To complement the multitude of ongoing programs to identify and
research $z<2$ galaxies, a next generation space telescope is 
essential to investigate the gas
which feeds and records the history of galaxy formation. 

\end{abstract}

\section{What is the Problem?}

In the next decade, numerous observational programs will identify and 
investigate overwhelming numbers of $z<2$ galaxies. These surveys will 
characterize the history of star formation, the evolution of galaxy
morphology, and the assembly of large-scale structure.  Altogether,
these efforts are likely to revolutionize our view of stars and
galaxies at $z<2$.  In contrast to these achievements, 
these observations will have minimal
impact on our understanding of the gas which feeds star formation.
At all redshift, gas is the major baryonic component of the universe
and, at $z>1$, probably the dominant baryonic component
of most galaxies. 
The principal challenge associated with examining the physics 
(e.g.\ metallicity,
density, ionization state, temperature) of this gas is that the
majority of diagnostics lie within the ultraviolet (UV) pass-band.
To cover this epoch, one requires a next generation space telescope.

With the advent of echelle spectrographs on 10m-class optical telescopes,
researchers have pursued quasar absorption line (QAL) studies at 
unprecedented levels in the early universe.  These observations have
revolutionized our understanding of the \lya forest 
(e.g.\ Miralda-Escud\'e et al.\ 1996; Rauch 1998), 
measured the baryonic density of the universe 
(Burles \& Tytler 1998; Rauch et al.\ 1997),
revealed metals in among the least overdense structures observed
(Tytler et al.\ 1995; Ellison et al.\ 2000), provided new insight into the
chemical enrichment history of the universe (Pettini et al.\ 1997; Prochaska \&
Wolfe 2000), and traced the velocity fields of protogalaxies 
(Prochaska \& Wolfe 1997).  
These studies have tremendous impact on our theoretical description of
gas in the early universe.  For example, aside from the CMB, comparisons of CDM
predictions with the \lya forest stand as one of the greatest successes
of this cosmological paradigm.  Furthermore, the majority of theorists 
at this workshop stressed one particular point: advances in 'gastrophysics'
are essential to addressing the next theoretical frontier.  
Empirical constraints on 'gastrophysics' can only be made through observations
of gas and QAL observations provide the most efficient avenue
of investigation.

In contrast with the $z=2-5$ universe, there are very few 
diagnostics of the IGM or extragalactic ISM at $z<2$.  This antithesis
is easily explained: the majority of physical diagnostics have 
observed wavelengths below 3000\AA\ at $z<1.5$.
In particular, the \lya transition (at 1215\AA, the reddest H\,I resonance
line) can only be observed at $z>1.6$ with optical spectrographs.  Without
coverage of this key transition, one cannot begin to 
address the preceding list of scientific inquiry.

An extragalactic observer like myself might be prone to nonchalantly
dismiss the $z<2$ universe.  After all, this represents only 25$\%$ of
the redshift path accessible to QAL analysis. 
This perspective, however, is horribly skewed.
The epoch spanning $z=[0,1.5]$ encompasses roughly 80$\%$ of the current
age of the universe.  If one considers the temporal evolution of any
quantity (e.g.\ chemical enrichment, galaxy clustering, luminosity
functions), then to overlook this epoch is to remain ignorant of the
universe.

In this brief proceeding -- written in support of a next generation UV 
telescope -- I will emphasize the preceding introduction through
a review of one major area of QAL research: the damped \lya systems (DLA).
These QAL systems are defined to have an H\,I surface density in excess
of $10^{20.3} \cm{-2}$ and they dominate the universal H\,I content
at all epochs following reionization.  At $z>2$, the DLA are believed
to be the progenitors of present-day galaxies (Kauffmann 1996; Steinmetz 2002)
and echelle observations of the sightlines penetrating these galaxies provide
detailed physical measurements of the protogalactic ISM.  
The measurements include chemical enrichment level, 
star formation rate, dust content, nucleosynthetic enrichment history,
as well as clues to the pressure, temperature, and ionization balance
of this multi-phase medium.
These observations are directly analogous to observations of the Galaxy,
LMC, and SMC carried out by HST and past UV telescopes.

This is the central thesis of this proceeding: current observations reveal
the physics of galactic gas locally and at $z>2$ but very rarely address
the $\approx 10$~Gyr in between.  To probe what amounts to $80\%$ of the
universe, one needs a next generation space telescope.
In the following, I will highlight some of the major results of research
on damped \lya systems, primarily those related to my own research with
optical echelle spectrographs.  To pursue these same research areas
at $z<2$ will require similar instrumentation in the UV pass-band.

\section{What We Would Like to Pursue at $z<2$}

\subsection{H\,I and Metal-Enrichment}

Damped \lya systems derive their name from the observed quantum mechanical
damping of the \lya transition relating to their very large 
H\,I column densities, $\N{HI}$.
Because the \lya profile is dominated by this damping, a standard fit
to the observed profile has two free parameters: (i) the centroid or $z_{abs}$;
and (ii) $\N{HI}$.  Accurate measures of $\N{HI}$ can be acquired
with modest resolution and S/N.  Wolfe and his collaborators initiated
surveys for these galaxies 20 years ago (e.g.\ Wolfe et al.\ 1986;
Storrie-Lombardi \& Wolfe 2000) and
the majority of research has been performed on 4m-class telescopes.
Figure~1 summarizes the principal result from these H\,I surveys:
the cosmic evolution of $\Omega_{DLA}$, the universal mass density of
neutral gas in units of the critical density.  Focus on the measurements
at $z<2$ which were derived from IUE data (solid triangle;
Lanzetta, Wolfe, \& Turnshek 1995)
and HST follow-up observations of a Mg\,II-selected sample 
(solid squares; Rao \& Turnshek 2000).  The uncertainties in each set of measurements
are very large (those are logarithmic error bars!) and one notes
a stark disagreement between the central values of the two surveys.
These uncertainties emphasize the current challenge of studying
H\,I gas at $z<2$ with existing UV spectrographs.
While COS+HST will enable a modest survey of DLA at $z<0.6$, the parameter
space $z = [0.6,1.7]$ will require a next generation space telescope.

\begin{figure}
\plotfiddle{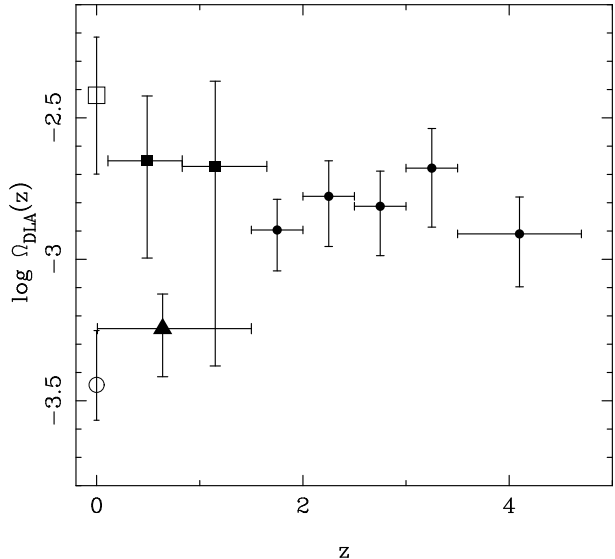}{2.7truein}{0}{60}{60}{-190}{-170}
\caption{Evolution in the cosmological H\,I density of the universe $\Omega_{DLA}$ 
in units of critical density as derived from DLA observations.
Note in particular, the large uncertainties in $\Omega_{DLA}$ at
$z<1.7$ (solid squares and triangle) and the large offset between 
these two UV surveys 
(IUE: Lanzetta, Wolfe, \& Turnshek 1995; HST: Rao \& Turnshek 2000).
These uncertainties highlight the inability of current UV spectrographs
to address this fundamental measure.  
}
\label{fig:HI}
\end{figure}

Aside from the H\,I content, the most basic measure of the galactic
ISM is metallicity.  Because of the large H\,I surface density of
DLA, ionization
corrections are small and measurements of low-ions like Fe$^+$, Si$^+$,
and Zn$^+$ yield accurate measures of the metallicity, i.e., 
[Zn/H]~$\approx$~[Zn$^+$/H$^0$].  The only serious
systematic error is dust
depletion; refractory elements like Fe and Si might be depleted from
the gas-phase such that Si$^+$/H$^0$ and Fe$^+$/H$^0$ are lower limits
to the true metallicity.  In general, the depletion levels of
the DLA are small (Pettini et al.\ 1997; Prochaska \& Wolfe 2002) and
the basic picture is well revealed by any of these elements at high $z$.
Metallicity observations of a large sample of DLA present
two main results: (1) an $\N{HI}$-weighted mean $<Z>$ which is the
cosmic mean metallicity of neutral gas;
and (2) metallicities for a set of galaxies which presumably span a 
large range of mass, morphology, and luminosity.  

\begin{figure}
\label{fig:metal}
\plotfiddle{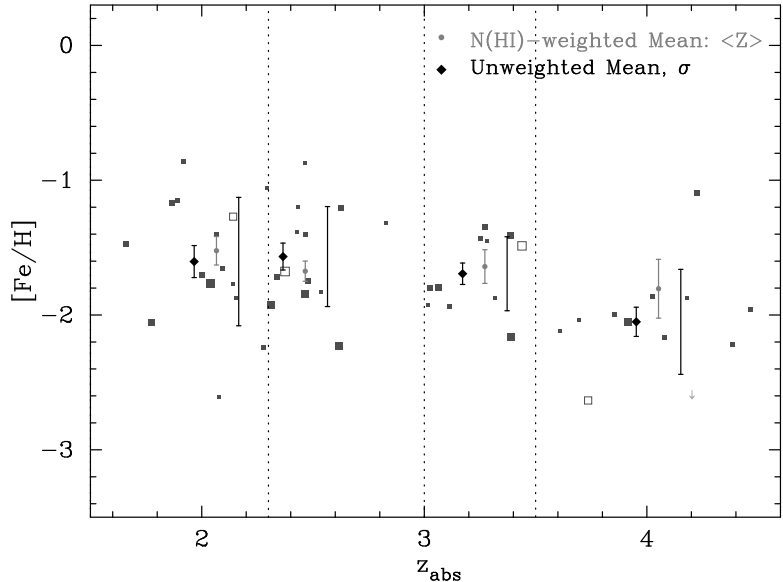}{2.7in}{270}{43}{43}{-160}{240}
\caption{Metallicity measurements for $\approx 50$~DLA at 
redshift $z \approx 2-4.5$. Overplotted are the H\,I-weighted
and unweighted means in several redshift bins.  
Interestingly, one observes minimal evolution in these mean metallicities.
Of particular interest, is resolving how the gas chemical enrichment
evolves from $z=2$ to today.
}
\end{figure}

Figure~2 presents over 50 metallicity measurements 
from $z \approx 2-4.5$ (Prochaska \& Wolfe 2002).
These are the principal results: (1) the mean metallicity (weighted or
unweighted) is significantly sub-solar;
(2) there is little evolution in the mean metallicity over this
redshift range with the possible exception of a modest
decrease at $z>3.5$; 
(3) no galaxy exhibits a metallicity lower than 1/1000 solar.
These optical observations constrain
models of chemical evolution at these epochs (e.g.\ Pei, Fall, \& Hauser 1999)
and give the first glimpse into metal production in the early universe. It
is crucial, however, to press to lower redshift.  
The time encompassed by the redshift interval $2 < z < 4.5$
pales in comparison with $z<2$.  Of immediate concern is to determine
how the mean metallicity rises to the enrichment level 
observed today.  

\begin{figure}
\plotfiddle{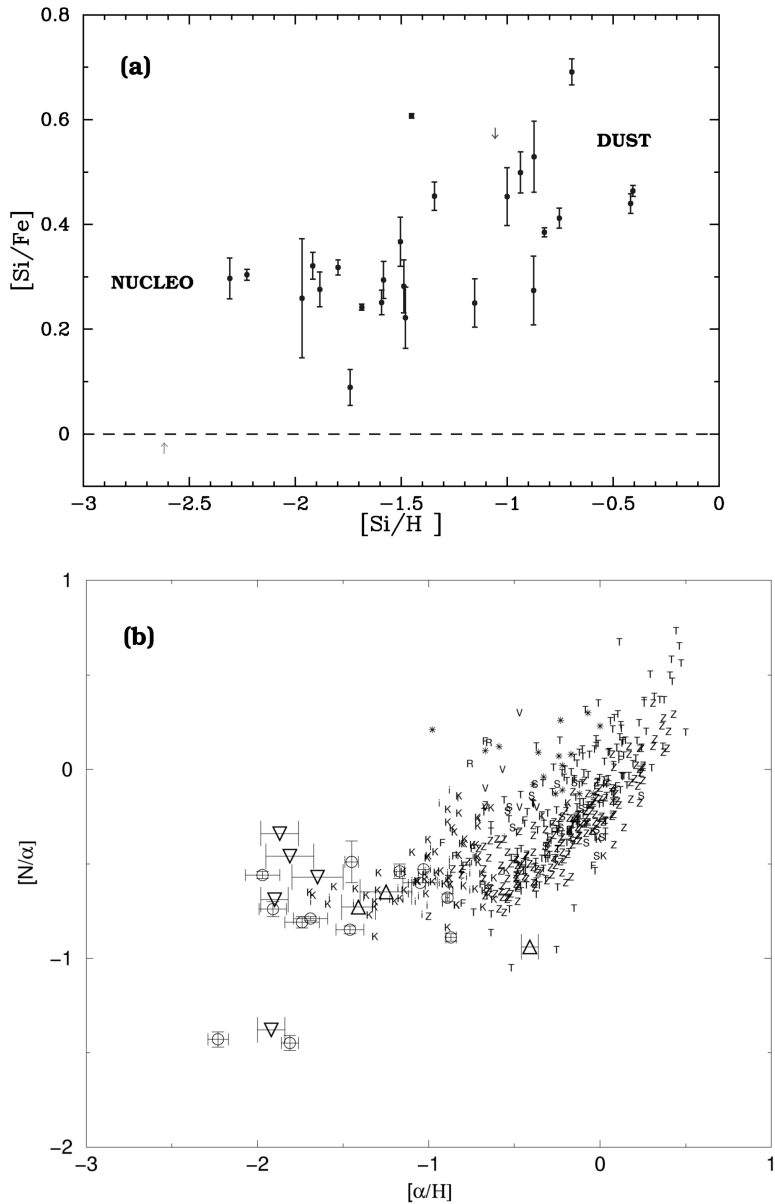}{6.5truein}{0}{80}{80}{-160}{0}
\caption{Relative abundances of (a) Si/Fe and (b) \nalph\ versus
Si/H and \alphH\ where $\alpha$ refers to either Si or S for the DLA.
The upper panel highlights the competing effects of nucleosynthesis
and dust depletion in interpreting gas-phase abundances.  We currently interpret
the plateau of [Si/Fe] at [Si/H]~$< -1.5$ as the result of Type~II SN
nucleosynthesis and are confident the rise in Si/Fe at [Si/H]~$>-1$
is associated with differential depletion (Prochaska \& Wolfe 2002).
UV observations would allow one to trace these two processes at $z<2$.
The lower panel compares DLA (circles and triangles; the latter are
upper/lower limits) against \nalph\ measurements for $z\sim 0$ H\,II
regions and stars (Prochaska et al.\ 2002; Henry, 
Edmunds, \& K{\"o}ppen 2000).   Although the majority of DLA lie
on the \nalph\ plateau observed in metal-poor H\,II regions,
a significant sub-sample is identified at \nalph~$< -1$.
This sub-sample is cautiously interpreted by Prochaska et al.\ (2002)
as evidence for a truncated or top-heavy IMF.  \nalph\ observations
at $z<2$ would help reveal the timescales of SF and the nature of the 
IMF at these epochs.
}
\label{fig:abund}
\end{figure}

\subsection{Relative Abundances: Dust and Nucleosynthesis}

High resolution ($R>30000$), high S/N ($>30$~res$^{-1}$), 
observations of the damped \lya systems enable detailed studies
of nucleosynthetic enrichment and dust properties in the early 
universe.  This level of data quality is crucial to achieving the
better than $10\%$ precision required by relative abundance studies.
Currently, there is an entire 'cottage industry' focused
on this area (Lu et al.\ 1996; Prochaska \& Wolfe 1999; Molaro et al.\ 2000;
Pettini et al.\ 2000; Ledoux, Bergeron, \& Petitjean 2002).  
Figure~3 presents two of the
principal results from our efforts: (a) [Si/Fe] and (b) [\nalph] 
measurements against [Si/H] metallicity. 

The super-solar Si/Fe ratios presented in panel~(a) highlight the
greatest obstacle
to interpreting relative abundance measurements from gas-phase
abundances: the competing effects of nucleosynthetic enrichment and
differential depletion.  In terms of dust depletion, 
one observes Si/Fe enhancements in depleted gas owing to the
differential depletion of these two refractory elements.
Regarding nucleosynthesis, Si/Fe enhancements suggest Type~II SN 
nucleosynthesis (e.g.\ Woosley \& Weaver 1995) whereas solar ratios
would imply Type~Ia SN enrichment patterns.  
Currently, we interpret the plateau
of Si/Fe values at low metallicity as the primary result of nucleosynthesis.
The mean enhancement matches the Galactic halo-star observations at
the same metallicity (e.g.\ McWilliam 1997) and it would be difficult to
understand why differential depletion would imply such a uniform enhancement.
In contrast, the rise in Si/Fe at [Si/H]~$>-1$ is highly suggestive
of differential depletion.  One expects a decrease in Si/Fe from
nucleosynthesis at higher metallicity
due to the increasing contribution from Type~Ia SN 
and larger depletion levels are sensible
in a higher metallicity ISM.
Investigating evolution in abundance ratios like these at $z<2$ 
would reveal
the detailed enrichment history of galaxies and the evolution of 
dust formation.

Overcoming this dust/nucleosynthesis degeneracy is among the most
active areas of DLA research.  One avenue is to focus on special pairs
of elements which are especially sensitive to only one effect.
Panel (b) is an excellent example of this; plotted are \nalph\ pairs
from a recent analysis by Prochaska et al.\ (2002).  For N, S, and Si
(the latter two are $\alpha$-elements), 
depletion effects are small and the results show
the nucleosynthetic history of N in the DLA.  For comparison, we
also plot [\nalph], [\alphH] pairs for $z \sim 0$ H\,II regions and
stars (see Henry, Edmunds, \& K{\"o}ppen 2000).  The majority of
DLA observations fall along the locus of local measurements, 
in particular the plateau of \nalph\ values at [Si/H]~$< -1$.
In contrast, a sub-sample of low metallicity DLA exhibit 
much lower \nalph\ values which Prochaska et al.\ (2002) interpret
these in terms of a truncated or top-heavy IMF.  These observations
have important implications for the processes of star formation in the
early universe and measurements at $z<2$ would assess the 
timescale of star formation in these galaxies and further elucidate
the nucleosynthesis of nitrogen. 

\begin{figure}
\plotfiddle{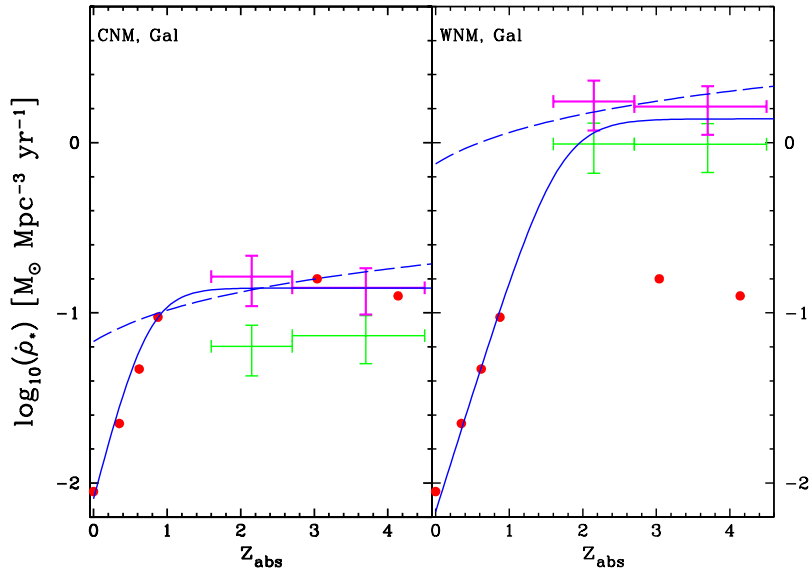}{2.7truein}{270}{40}{40}{-170}{223}
\caption{Star formation histories derived from the damped \lya systems
(points with error bars) compared against similar measures from the 
LBG (Steidel et al.\ 1996) and low $z$ UV-samples (Treyer et al.\ 1998).
The DLA observations provide a measure of the SF history in a sample
of galaxies unbiased to luminosity.  
}
\label{fig:sfr}
\end{figure}

\subsection{Star Formation}

Through observations of the C\,II$^*$~1335 fine-structure transition,
it is possible to infer the star formation rate per unit area (SFR/area)
within DLA. 
In the Milky Way, the dominant cooling
mechanism of the ISM is [CII] 158~\micron\ emission from collisionally
excited C$^+$ (Wright et al.\ 1991).  Assuming steady-state
equilibrium {\it and} that star formation dominates the ISM heating,
one can infer the SFR/area from observations of $\N{CII^*}/\N{HI}$
(Wolfe, Prochaska, \& Gawiser 2002).
In practice, the technique is complicated; one must self-consistently
solve for the two-phase medium of the ISM and address several
dust-related issues.   Nevertheless, the method allows for an 
assessment of the SFR/area of the DLA and thereby the SFR density $\dot \rho_*$
of these protogalaxies.  Figure~4 presents a preliminary
picture of the SF history of the DLA assuming a two-phase medium
with cooling dominated by the (a) CNM and (b) WNM.  
The two sets of data points with error bars in each panel are the
DLA measurements for two assumptions of dust depletion, the solid
points are $\dot \rho_*, z$ pairs from other surveys
(Steidel et al.\ 1996; Treyer et al.\ 1998), 
and the solid and dashed lines are assumed fits to the DLA data.
Because limits on integrated light (e.g.\ Bernstein, Freedman, \& Madore 2002)
rule out the WNM solution, focus on the rates for the CNM.
Our observations suggest $\dot \rho_*, z$ values consistent with or 
somewhat lower than those inferred from LBG observations.  
Understanding the connection between these two galactic populations
will be very important for resolving the history of galaxy formation. 
One can apply these same techniques at $z < 2$ with UV spectroscopy.
The results would offer a complimentary assessment of the
SFR from emission-line and integrated light techniques.
These SF diagnostics are critical for deciphering the build-up of
galactic structure and metal enrichment in our universe.

\subsection{Velocity Fields, $\alpha$, etc.}

The space allotted for this proceeding precludes a full discussion
of the impact of a next generation UV telescope on even DLA research. 
To list a few areas I will neglect:
(1) velocity fields: measurements of the gas dynamics in DLA yield
mass estimates in a fashion which avoids the 
luminosity-bias of most traditional approaches (Prochaska \& Wolfe 1997). 
Gas dynamics are sensitive to non-gravitational motions (e.g.\ SN feedback),
however, and interpretation may not be
straightforward.  Nevertheless, comparisons of QAL velocity widths
with stellar measures would lend further insight into measuring
galactic masses;
(2) H$_2$: With far-UV spectroscopy, one can examine molecular hydrogen
in the DLA (e.g.\ Bechtold 2002) and thereby examine the 
gas which serves as the precursor to star formation.  Additionally, one
gains insight into dust formation and several physical characteristics
of the gas (e.g.\ temperature, UV radiation field);
(3) $\alpha$: observations of DLA at $z>2$ have enabled an analysis
of the evolution of the fine-structure constant (e.g.\ Murphy et al.\ 2001)
and have implied possible variations in this fundamental physical constant.
UV observations would allow us to perform a similar inquiry at $z<2$.

\section{What We Want or What We Need}

The resolution and S/N needed to examine the DLA or perform
QAL studies in general is dependent on the specific research area.
Nevertheless, we can lean on our extensive experience with high $z$
QAL observations using optical spectrographs.
For the majority of scientific applications, $R=30000$ is a bare
minimum.  Only at this resolution can one confidently distinguish
\lya clouds from metal-lines in the \lya forest, obtain abundance
measurements to greater than 0.1~dex precision, resolve velocity
fields, and investigate a multi-phase medium.
Note this value is 50$\%$ greater than the $R=20000$ of COS+HST.
I fear this instrument will have limited impact on several important
research areas, e.g., relative chemical abundances, kinematic 
characteristics.

Regarding S/N, a good lower limit is 30 per resolution element
(i.e.\ 15 pix$^{-1}$ for
4 pixel sampling).  At this level, one can carefully address systematic
errors like continuum placement and also analyse a wealth of very
powerful absorption line diagnostics (e.g.\ Si\,II~1808, C\,IV~1550,
O\,VI~1030).  As always, one gains by achieving higher S/N and many
applications would depend on higher sensitivity.  

Now, consider wavelength coverage.  For the majority of QAL research,
coverage from \lya (1215\AA) to 2000~\AA\ rest-frame is essential.
Many projects (e.g.\ D/H measurements, photoionization assessment, H$_2$
observations) need coverage down to 1000 or even 900\AA\ rest-frame.
It is my opinion that an instrument providing coverage from 1100--3200\AA\
would be ideal with 1200--3000\AA\ acceptable.

Finally, and perhaps most importantly, observing power.  To
allow observations of a large enough sample of QSO's at $z<2$, 
a UV telescope must achieve the
above resolution, S/N, and wavelength coverage for a $V \approx 18$ QSO
in a reasonable exposure time ($<10$~hr).  This would provide enough
targets to examine the physical conditions at similar levels as
achieved at $z>2$.
Taking the predicted performance of COS+HST
as a starting point, the next generation space telescope would need
a 5--10$\times$ improvement over COS+HST in its highest throughput 
pass-band (1100--2000\AA) and an increase of $50\times$ over COS+HST
at $\lambda > 2000$\AA.  Although improvements in coatings and 
instrument design could make-up $3\times$ of the difference 
(probably over $10\times$ at $\lambda > 2000$\AA), a larger aperture
telescope is unavoidable.  After attending this workshop,
I feel a 4m aperture might just satisfy the constraints I have laid out
and a $\geq 6$m aperture would certainly meet the demands.
Obviously, this will require terrific funding from NASA
and, therefore, considerable support from the astronomical community.
I am convinced, however, that the value of examining the gas physics
of the universe at $z<2$, an effort which complements numerous
observational programs within NASA's ORIGINS and SEU Themes,
is worth the expense.

\acknowledgements

I would like to thank S. Rao and A. Wolfe who provided the H\,I and
C\,II$^*$ figures.  I also wish to apologize to the many
folk in the DLA and ISM community who I failed to reference in this
brief proceeding.

\end{document}